\documentclass[lettersize,journal]{IEEEtran}
\usepackage[utf8]{inputenc}
\usepackage{amsthm}
\usepackage{amssymb}
\usepackage{amsmath}
\usepackage{graphicx}
\usepackage{subfig}
\usepackage{lineno}
\usepackage{xcolor}

\newcommand{\const }[0]{\operatorname{const}}

\newcommand\largefrac[2]{\frac{\displaystyle #1}{\displaystyle #2}}
\allowdisplaybreaks
\begin{document}

\title{Dynamic Virtual Inertia and Damping Control for Zero-Inertia Grids}
\author{Oleg O. Khamisov and Stepan P. Vasilev
\thanks{Oleg O. Khamisov and Stepan P. Vasilev are with Center for Energy Science \& Technology, Skolkovo Institute of Science \& Technology, Moscow, Russian Federation (e-mail: o.khamisov@skoltech.ru, stepan.vasilev@skoltech.ru).}
\thanks{Stepan P. Vasilev is with Sber AI Lab, Moscow 117312, Russia.}}
\maketitle

\begin{abstract}
In this paper virtual synchronous generation (VSG) approach is investigated in application to low- and zero-inertia grids operated by grid-forming (GFM) inverters. The key idea here is to introduce dynamic inertia and damping constants in order to keep power gird stable during different types of faults, islanding or large power balance oscillations. In order to achieve such robustness, we introduce frequency and phase angle shift functions to VSG along with dynamics virtual generator parameters. The stability of such approach is theoretically proven and theoretical results are supported by detailed case studies in RTDS (Real-Time Digital Simulator) NovaCor 1.0 with GFM inverters dynamics simulated with 1-3 microseconds timestep using two-level universal inverter model. Case studies include all aforementioned types of faults and demonstrate increased power grid robustness and survivability in comparison with traditional synchronous generation of comparable size.
\end{abstract}


\section*{Nomenclature}
\addcontentsline{toc}{section}{Nomenclature}
\begin{IEEEdescription}[\IEEEusemathlabelsep\IEEEsetlabelwidth{$V_1,V_2,V_3$}]
{\color{black} 
\item[\textit{Abbreviations}]
\item[]
\item[AGC] Automatic Generation Control.
\item[IBR] Inverter-Based Resources.
\item[PLL] Phase-Locked Loop.
\item[RoCoF] Rate of Change of Frequency.
\item[EMT] Electro-magnetic Transient.
\item[GFM] Grid-Forming.
\item[SG] Synchronous Generator.
\item[VSG] Virtual Synchronous Generator.
\item[PSS] Power System Stabilizer.
\item[]
\item[\textit{Variables and parameters}]
\item[]
\item[$\omega_S$]  Power grid frequency deviation.
\item[$\theta_S$]  Power grid phase angle deviation.
\item[$P_G$] Aggregated power grid active power generation.
\item[$P_L$] Aggregated power grid active power consumption.
\item[$V_S$] Aggregated power grid voltage magnitude.
\item[$G$] Conductance of a line connecting inverter to power grid.
\item[$B$] Susceptance of a line connecting inverter to power grid.
\item[$M_S$] Aggregated power grid inertia.
\item[$D_S$] Aggregated power grid dumping.

\item[$\omega_{ref}$] Reference frequency.
\item[$\omega$] Inverter VSG frequency deviation.
\item[$\omega_{shift}$] Inverter VSG frequency deviation shift.
\item[$\omega_{set}$] Inverter VSG frequency.

\item[$\theta$] Inverter VSG phase angle deviation.
\item[$\theta_{shift}$] Inverter VSG phase angle deviation shift.
\item[$\theta_{set}$] Inverter VSG phase angle.

\item[$P_{out}$] Inverter terminal active power.
\item[$P_{ref}$] Inverter reference active power.
\item[$P_{max}$] Inverter maximal active power.
\item[$P_{min}$] Inverter minimal active power.
\item[$\gamma$] Filtered active power limits penalty.
\item[$\tau_\gamma$] Filter characteristic time for $\gamma$.

\item[$\tau_\omega$] Inverter VSG characteristic time.
\item[$\alpha$] Inverter VSG inverted dumping.
\item[$\alpha_{min}$] Inverted dumping lower limit.
\item[$c_\alpha$] Inverted dumping shift coefficient.
\item[$c_\theta$] Inverter phase angle shift coefficient.
\item[$c_\omega$] Inverter frequency shift coefficient.
\item[$A$] Linearized matrix of VSG dynamics.
\item[$B$] Matrix with non-diagonal elements of $A$.

\item[$V_{out}$] Inverter terminal voltage magnitude.
\item[$V_{max}$] Inverter maximal voltage magnitude.
\item[$V_{min}$] Inverter minimal voltage magnitude.
\item[$Q_{out}$] Inverter terminal reactive power.


\item[$X$] SG step-up transformer reactance.
\item[$T''_q$] SG open circuit q-axis time constant
\item[$X'_q$] SG q-axis transient reactance.
\item[$X''_q$] SG q-axis subtransient reactance.
\item[$\mathbb{R}$] Set of real numbers

\item[]
\item[\textit{Functions}]
\item[]
\item[$f(\theta,t)$] Inverter active power output at phase angle $\theta$.
\item[$f_{ref}(\theta,t)$] Inverter active power output deviation from reference power $P_{ref}$ at phase angel $\theta$.
\item[$f_{max}(\theta,t)$] Inverter active power output deviation from maximal power $P_{max}$ at phase angle $\theta$.
}
\end{IEEEdescription}

\section{Introduction}\label{s:intro}
Integration of inverter-based resources (IBRs) into modern power grids drastically affects dynamic performances of power systems. As a result, reduced power system inertia requires faster response from frequency control algorithms. Majority of IBRs operate in grid-following control (GFL) strategy with voltage and frequency explicitly measured in the point of interconnection with phase-locked loop (PLL) \cite{kroposki2017, milano2018}. This strategy is limited to grids with a significant amount of synchronous generation capable of providing adequate reference points. High penetration of GFL-based inverters is impractical and can cause instability \cite{wang2018, nerc2017, lin2017}. As renewable energy integration increases, the focus has shifted to grid-forming (GFM) control. In this type of control IBRs create local voltage phasors and manage synchronization through frequency droop \cite{sajadi2022}, allowing direct control over output frequency and active power, and similarly over voltage and reactive power \cite{nerc2021}. Previous studies suggest that GFMs can address frequency stability issues in low inertia systems \cite{lin2020, markovic2021}. Previous studies have focused on small-signal stability in systems with GFMs, often concluding the necessity of some SG presence for stability while overlooking new frequency regulation possibilities with GFMs \cite{markovic2021}. Simultaneously, the concept of grid-forming (GFM) control ideally operates without synchronous generation. However, its practical implementation presents challenges, as illustrated in several studies \cite{alipoor2015, wu2020, wu2019}. Generally, grid-following inverters utilize a PLL for synchronization \cite{song2017}, whereas GFM methods are thought not to use PLL, or only for initial synchronization \cite{wu2019}. Nonetheless, PLL-based frequency and angle measurements can be used to enhance GFM control \cite{sun2021}. For parallel operation, all GFM units require synchronization, and PLL usage does not inherently conflict with GFM functionality. Power grids can operate solely with inverters using an advanced current-controlled scheme \cite{quan2020}. Therefore, it is not entirely accurate to claim that GFMs never use PLL \cite{unruh}. Other concept of GFM control is virtual synchronous generation. Idea of virtual synchronous generation was firstly presented in \cite{beck2007} and is aimed to remedy the aforementioned instabilities in zero-inertia grids. The possibility of 100\% GFM generation was presented in \cite{kenyon2021} with numerical simulation of corresponding EMT. Later, in \cite{tayyebi2020} the authors demonstrated SG operation together with GFM. Particularly in scenarios where GFMs augment SG-driven inertial responses. Other research highlights the damping contributions of droop-controlled GFMs to frequency dynamics, typically within SG-dominated frameworks \cite{lasseter2020}. The development of control designs in this area has complications due to high standard for verifiable simulations. As it was shown in \cite{KENYON2023108789}, standard timestep of quarter electrical cycle is usually insufficient, and detailed simulations of inverter dynamics should be done in the order of microseconds.

The main contributions of this work can be separated into 3 following items:
\begin{enumerate}
    \item VSG control system that dynamically adjusts system inertia and dumping in order to maximize robustness of zero-inertia power grid.
    \item Simulation of 100\% IBR 9 bus power grid \cite{Sauer1998} with universal converter models operating at 1-3 microseconds timestep \cite{en13164036} implemented in RTDS NovaCor 1.0.
    \item Detailed robustness analysis of the proposed control system, including usage of IBRs only and IBRs with SG under different system faults.
\end{enumerate}

The paper is organised in the following way. Section \ref{s:control} contains development of the VSG controller with dynamic inertia and dumping coefficients. Section \ref{s:setup} contains RTDS model of 9-bus system with possibility to switch between IBRs and SGs. Section \ref{s:tests} is dedicated to system robustness tests. Finally section \ref{s:conclusion} is conclusion of the work.
\section{GFM controls}\label{s:control}
\begin{figure}
    \centering
    \includegraphics[width=0.6\linewidth]{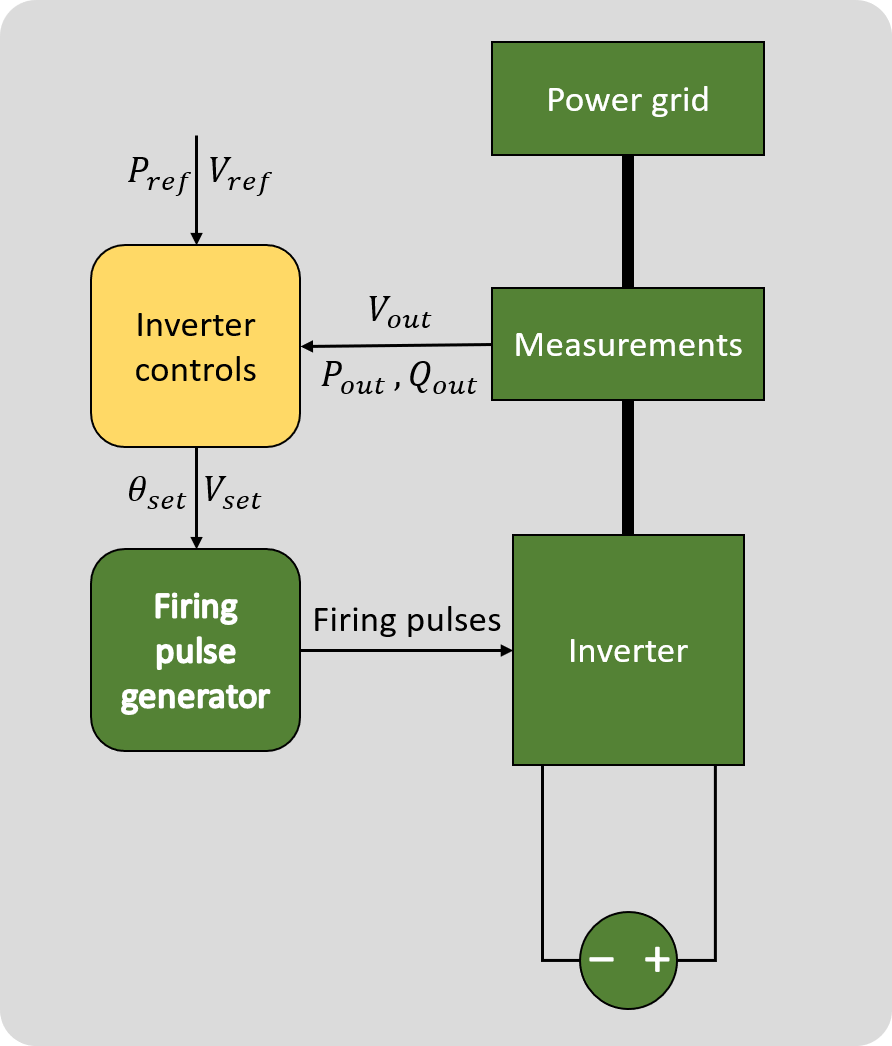}
    \caption{Inverter setup}
    \label{f:setup}
\end{figure}

This section is dedicated to the development of grid-forming inverter controls. It is assumed, that developed controller resolves measurements of voltage magnitude, actual active, reactive powers and reference values of voltage and active power (which are send by higher control loops, i.e. AGC or PSS). The controller outputs phase angle and voltage magnitude that are then used in firing pulse generator of the inverter (Fig \ref{f:setup}). It is assumed that actual active and reactive powers ($P_{out}$ and $Q_{out}$) together with voltage magnitude $V_{out}$ are control inputs together with active power and voltage magnitude reference values ($P_{ref}$ and $V_{ref}$). Then control at each point in time generates desired voltage magnitude and phase angle ($V_{set}$ and $\theta_{set}$) for the firing pulse generator, which, in its turn sends pulses to the inverter.

In order to improve control performance in comparison with other VSG techniques, we utilize the idea of frequency and phase angle shifts during EMT, which are impossible to introduce in SG. Their purpose is to allow the system to have large virtual inertia without violation of current limitations.

Note that active power output of inverter is defined by power flow equations
\begin{equation}\label{e:pf}
    P_{out} = V_{out}V_S\left(G\cos(\theta_{out}-\theta_S)+B\sin(\theta_{out}-\theta_S)\right),
\end{equation}
where $G$ and $B$ are conductance and susceptance respectively, $V_S$ and $\theta_S$ are power system voltage magnitude and phase angle respectively.
Since parameterise in power flows equations can vary significantly from grid to grid, for the control derivation we introduce function $f:\mathbb{R}^2\rightarrow\mathbb{R}$ such that 
\begin{equation}
    P_{out} = f(\theta_{out},t).
\end{equation}
Assuming, that inverter dynamics affect insignificantly grid variables during EMT, $V_S$ and $\theta_S$ depend on time only. Thus, its derivative in $\theta$ is given by
\begin{equation}
    \frac{\partial f}{\partial\theta_{out}} = V_{out}V_s\left(B\cos(\theta_{out}-\theta_S)-G\sin(\theta_{out}-\theta_S)\right).
\end{equation}
As a result, for limited difference between $\theta_{out}$ and $\theta_S$, the derivative is positive and function $f$ is monotonous in $\theta_{out}$. Moreover, for transmission grids $B\gg G$ and monotonicity is kept for angle differences close to $\pi/2$. Introduction of function $f$ will allow to derive inverter control with power grid being a black box with the monotonicity of $f$ property.

\begin{figure*}
    \centering
    \includegraphics[width=1\linewidth]{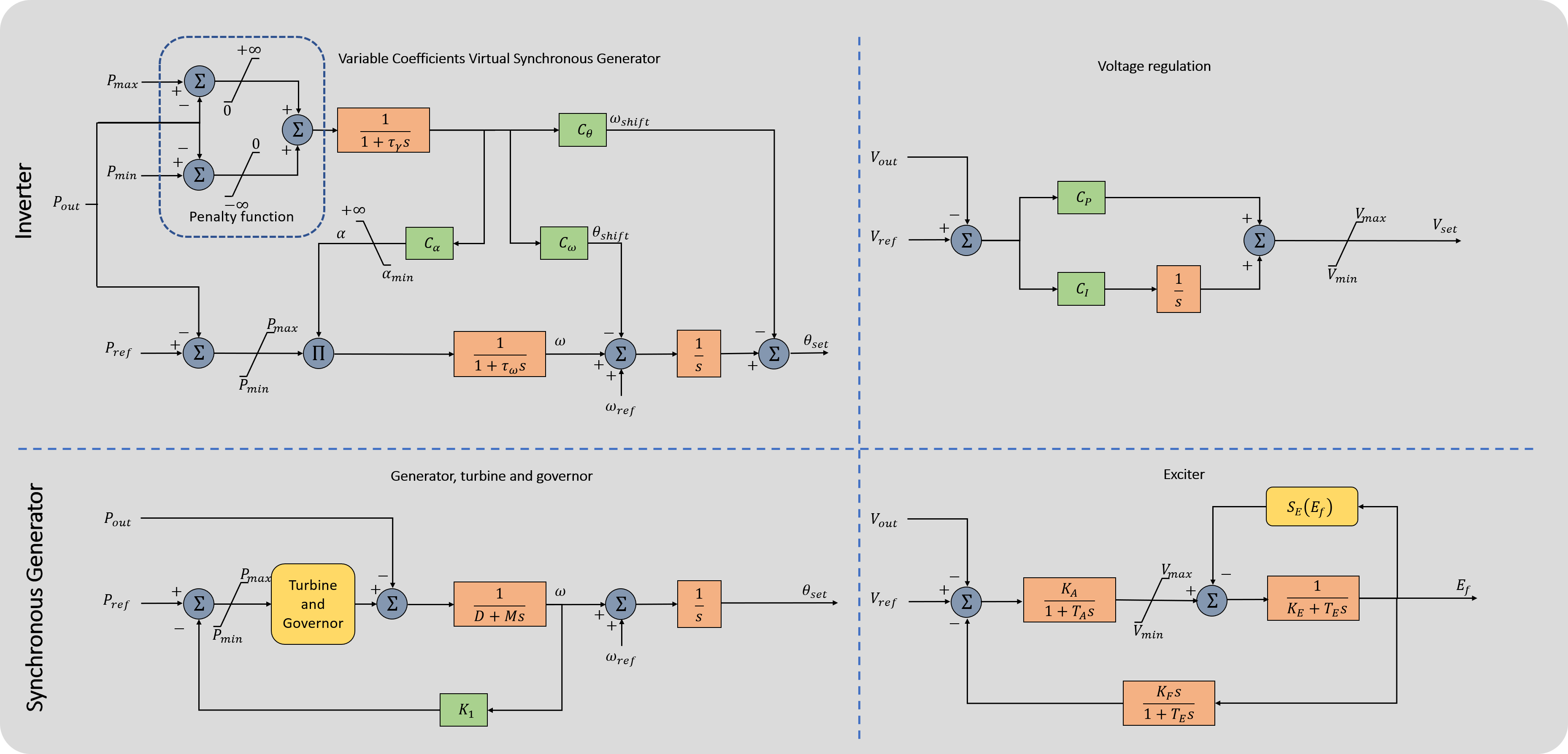}
    \caption{Comparison of inverter and SG controls.}
    \label{fig:controller}
\end{figure*}

Further control development will be separated into two subsections: control block for $\theta_{set}$ and control block for $V_{set}$. General control logic and its comparison with SG is given in the Fig. \ref{fig:controller}. Components for SGs are taken form \cite{MBB}. Detailed description of the developed control is given in the following subsections.

\subsection{VSG controller}
The goal of VSG is to minimize frequency oscillations using second order synchronous machine dynamics, while keeping voltage within the acceptable limits. The difficulty of calculating optimal control response lays in the detailed dynamic model of power system. It is highly nonlinear and in general is non-observable and non-controllable from optimal control perspective. Thus, in this work we use a control design approach that consist of multiple simplified steps which are later verified by a detailed case study. These steps are the following:
\begin{enumerate}
    \item Calculate optimal inverter power output assuming that system dynamics are aggregated to a single bus and inverter dynamics are instant;
    \item Develop virtual inertia and dumping responses that will achieve the desired power output;
    \item Adjust controller to account for inverter dynamics, delays in control response and measurement inaccuracies.
\end{enumerate}

Let us begin with simplified problem. It is formulated as minimization of aggregated frequency deviation over classical generator model dynamics. At this step we assume, that one IBR is controlled. All other generation consists of SG and IBRs with VSG. Dynamics of all IBRs are ignored. As a result, problem statement has the following form:
\begin{subequations}\label{e:p}
\begin{equation}
    \min\frac{1}{2}\int_0^T\omega_S^2(t)dt,
\end{equation}
\begin{equation}\label{e:p:sg}
    M_S\dot\omega_S = -D_S\omega_S + P_G - P_L + P_{out},
\end{equation}
\begin{equation}
    P_{out}\in[P_{min},P_{max}].
\end{equation}
\end{subequations}
Here lower index "$S$" is used to emphasise that these variables are power system variables and not inverter variables. Variable $\omega_S$ is a frequency deviation from reference point, $\theta_S$ is a phase angle, $P_G$ is generation of SGs and IBRs that are not controlled within this problem, $P_{out}$ is power output of controlled IBR. $M_S$ and $D_S$ are system inertia and dumping coefficients respectively. For
\begin{equation}\label{e:const}
    P_G - P_L\equiv\const
\end{equation}
this problem can be solved via direct application of Pontryagin's Maximum Principle, as it was shown in \cite{KVI} The solution is given by
\begin{equation}\label{e:sol}
    P_{out}(t) = \left\{\begin{array}{ll}
        P_{min}, & \omega_S(t)>0, \\
        P_{max}, & \omega_S(t)<0, \\
        P_G - P_L, & \omega_S(t) = 0.
    \end{array}\right.
\end{equation}
While it is possible to obtain explicit solution \eqref{e:sol}, problem \eqref{e:p} has very strong assumption: grid topology is ignored. Moreover, it requires instant bang-bang changes in inverter power output, which are not possible to implement even with the fast inverter dynamics.  Thus, we introduce VSG type of inverter control with the idea of dynamicall adjustment of its parameters to achieve inverter response close to ideal one \eqref{e:sol}. Let us introduce VSG models:
\begin{subequations}\label{e:vsg}
\begin{equation}\label{e:vsg:omega}
    \tau_\omega\dot\omega = -\omega + \alpha(P_{ref} - P_{out}),
\end{equation}
\begin{equation}
    \dot\theta = \omega,
\end{equation}
\begin{equation}
    P_{out} = f(\theta,t).
\end{equation}
\end{subequations}
Here $P_{ref}$ is active power reference, obtained from outer control loop (i.e. Automatic Generation Control) and is not a part of VSG or Primary regulation control loop. For the simplicity of further derivation, VSG equation \eqref{e:vsg:omega} differs from actual generator swing equation \eqref{e:p:sg}. This is done to analyse case, when $\alpha = 0$, which is not possible to simulate using equation \eqref{e:p:sg}. Nevertheless, equation \eqref{e:vsg:omega} has interpretation as a generator swing equation for $\alpha\ne0$ with virtual inertia and dumping coefficients are given by $\tau_\omega/\alpha$ and $1/\alpha$ respectively.

Now let us choose $\tau_\omega$ and $\alpha$ in a way that allows $P_{out}$ converge to the form \eqref{e:const}.
If system is in sliding mode  we have $P_{out} = P_G - P_L$ and $\omega(t) = 0$. Thus, $\alpha = 0$ and $\tau_\omega>0$.
For the case, when control is not in sliding mode, without loss of generality it is assumed, that power grid is in power deficit and frequency is below reference value. Due to bang-bang nature of the control \eqref{e:sol}, transition to maximum power output includes discontinuity over $\theta$  in the system \eqref{e:vsg}. However, $\theta$ is a solution of differential equation and is continues. In order to approximate discontinuity it is necessary to take $\dot\theta\rightarrow\infty$ to increase inverter power output, which yields $\omega\rightarrow\infty$ and $M\rightarrow0$. In practice, this means transition to a zero inertia system, with step-changes in the phase angle which is highly unstable. Instead, within this work, we propose to keep system inertia significantly large and introduce limited shift to output frequency and phase angle as is shown below:
\begin{subequations}\label{e:vsg_mod}
\begin{equation}\label{e:vsg_mod:omega}
    \tau_\omega\dot\omega = -\omega + \alpha(P_{ref} - P_{out}),
\end{equation}    
\begin{equation}
    \dot\theta = \omega - \omega^{shift},
\end{equation}
\begin{equation}
    P_{out} = f(\theta-\theta^{shift},t),
\end{equation}
\end{subequations}
where $\omega^{shift}$ and $\theta^{shift}$ depend linearly on difference between maximal power output and actual power output. To avoid algebraic control loop and to filter measurement noise, we introduce auxiliary variable $\gamma$, which converges to linear penalty for violating active power limits as is shown below:
\begin{subequations}\label{e:gamma_sys}
\begin{equation}
    \tau_\gamma\dot\gamma=-\gamma+\max(0,P_{out}-P_{max}),\;\tau_\gamma>0,
\end{equation}
\begin{equation}
    \theta_{shift} = c_\theta\gamma,\;c_\theta>0,
\end{equation}
\begin{equation}
    \omega_{shift} = c_\omega\gamma,c_\omega>0.
\end{equation}
\end{subequations}
Here positive constants $c_\theta$ and $c_\omega$ are chosen to ensure stability of VSG differential equations and will be discussed later. Additionally, in the \eqref{e:vsg_mod:omega} dependence on $P_{ref}$ is present. Since it is the only inhomogenity with respect to $\omega$ we modify it by assigning
\begin{equation}
    \alpha = c_\alpha\gamma,\;c_\alpha>0.
\end{equation}
Let us now define constants $c_\omega, c_\theta$ and $c_\alpha$. Recall, that function $f$ is monotonous over $\theta$ since we assume. Thus, it is possible to introduce two monotonous functions:
\begin{equation}
    f_{ref}(\theta,t) = f(\theta,t) - P_{ref},
\end{equation}
\begin{equation}
    f_{max}(\theta,t) = f(\theta,t) - P_{max}.
\end{equation}
This allows us to unite systems \eqref{e:vsg_mod} and \eqref{e:gamma_sys} into one system of differential equations (here and further it is assumed that $P_{out}\ge P_{max}$):
\begin{subequations}\label{e:vsg_fin}
\begin{equation}\label{e:vsg_fin:omega}
    \tau_\omega\dot\omega = -\omega - c_\alpha\gamma f_{ref}(\theta - c_\theta\gamma,t),
\end{equation}    
\begin{equation}\label{e:vsg_fin:theta}
    \dot\theta=\omega-c_\omega\gamma,
\end{equation}
\begin{equation}\label{e:vsg_fin:gamma}
    \tau_\gamma\dot\gamma = -\gamma + f_{max}(\theta - c_\theta\gamma,t).
\end{equation}
\end{subequations}
Firstly, let us consider stationary point of this system. After excluding $\omega$ via substitution from \eqref{e:vsg_fin:omega} into \eqref{e:vsg_fin:theta} we get the following stationary point equations:
\begin{subequations}
\begin{equation}
    \gamma\left(c_\alpha f_{ref}(\theta-c_\theta\gamma,t) + c_\omega\right) = 0,
\end{equation}    
\begin{equation}
    \gamma - f_{max}(\theta - c_\theta\gamma,t) = 0.
\end{equation}
\end{subequations}
If $\gamma\ne0$. Then $c_\alpha f_{ref}(\theta - c_\theta\gamma)+c_\omega = 0$ and $\gamma>0$, since it is assumed, that $P_{out}\ge P_{max}$. Additionally, $P_{max}\ge P_{ref}$ and $f_{ref}(\theta - c_\theta\gamma,t)\ge f_{max}(\theta - c_\theta\gamma)$. As a result
\begin{equation}
\begin{gathered}
    c_\alpha f_{ref}(\theta - c_\theta\gamma,t)+c_\omega\ge c_\alpha f_{max}(\theta - c_\theta\gamma,t)+c_\omega =\\= c_\alpha\gamma+c_\omega>0.
\end{gathered}
\end{equation}
This contradiction leaves only one solution $\gamma = 0$ and $f_{max}(\theta-c_\theta\gamma,t)=0$; thus, $P_{out} = P_{max}$ which is the desirable power output.

System \eqref{e:vsg_fin} is nonlinear. Thus, its stability will be analyzed via linearization. The corresponding system matrix at the point $(\omega_0,\theta_0,\gamma_0,t_0)$ has form
\begin{equation}
    A = \left(\begin{array}{ccc}
        -\largefrac{1}{\tau_\omega} & -\largefrac{1}{\tau_\omega}c_\theta\gamma_0\nabla f_{ref}^0 & -\largefrac{c_\theta}{\tau_\omega}(f_{ref}^0+c_\omega\nabla f_{ref}^0)\\
        1 & 0 & -c_\omega\\
        0 & \largefrac{1}{\tau_\gamma}\gamma_0\nabla f_{max}^0 & -\largefrac{1}{\tau_\gamma}(1+c_\theta\nabla f^{max}_0)
    \end{array}\right).
\end{equation}
where
\begin{equation}
    f_{ref}^0 = f(\theta_0-c_\theta\gamma_0,t_0)
\end{equation}
and
\begin{equation}
    \nabla f_{ref}^0 = \left.\frac{\partial f_{ref}}{\partial\theta}\right|_{(\theta,t) = (\theta_0-c_\theta\gamma_0,t_0)}.
\end{equation}
Matrix $A$ can be represented as a sum of diagonal negative semi-definite and matrix
\begin{equation}
    B = \left(\begin{array}{ccc}
       0 & -\largefrac{1}{\tau_\omega}c_\theta\gamma_0\nabla f_{ref}^0 & -\largefrac{c_\theta}{\tau_\omega}(f_{ref}^0+c_\omega\nabla f_{ref}^0)\\
        1 & 0 & -c_\omega\\
        0 & \largefrac{1}{\tau_\gamma}\gamma_0\nabla f_{max}^0 & 0
    \end{array}\right),
\end{equation}
Its eigenvalues can be found via standard formula
\begin{equation}\label{e:detb}
\begin{gathered}
    \det(B-I\lambda) = \lambda\Bigg(\lambda^2+\largefrac{1}{\tau_\omega}c_\theta\gamma_0\nabla f_{ref}^0+\\
    +\largefrac{1}{\tau_\gamma}\gamma_0\nabla f_{max}^0c_\omega\Bigg) = 0.
\end{gathered}
\end{equation}
Note, that $\gamma$ is non-negative by definition \eqref{e:vsg_fin:gamma}. Thus, $\gamma_0\ge0$. Additionally, $\nabla f^0_{ref}$ and $\nabla f^0_{max}$ are positive due to monotonicity and  all constants $\tau_\omega,\tau_\gamma,c_\theta$ and $c_\omega$ are positive. Second summand in the brackets in \eqref{e:detb} is non-negative, all roots $\lambda$ of the equation \eqref{e:detb} have real parts equal 0 and matrix $B$ is negative semi-definite. Finally, matrix $A$ also is negative semi-definite as a sum of two negative semi-definite matrices. Finally, if constants $c_\omega, c_\theta$ and $c_\alpha$ are chosen so that 
\begin{equation}\label{e:deta}
    \det A\ne0,
\end{equation}
system \eqref{e:vsg_fin} is asymptotically stable and corresponding power output converges to $P_{max}$.

Repeating the same approach for the lower limit $P_{min}$ is identical to $P_{max}$. This statement finalizes controller derivation with a small practical adjustment. Finally, in order for the secondary frequency control regulation to operate normally, it is necessary to keep small frequency deviation even during sliding mode and VSG with dynamical inertia and dumping is represented by the following system of equations:
\begin{subequations}
\begin{equation}
    \tau_\omega\dot\omega = - \omega + c_\alpha\gamma(P_{ref} - P_{out}),
\end{equation}
\begin{equation}
    \dot\theta = \omega - c_\omega\gamma,
\end{equation}
\begin{equation}
\begin{gathered}
    \tau_\gamma\dot\gamma = -\gamma + \max\{0,P_{out} - P_{max}\}+\\+\min\{0,P_{out} - P_{min}\},
\end{gathered}
\end{equation}
\begin{equation}
    \theta^{out} = \theta - c_\theta\gamma + \omega_{ref}t.
\end{equation}
\end{subequations}
The exact choice of the constants $\tau_\omega,\tau_\gamma,c_\omega, c_\theta$ and $c_\alpha$ is empirical with respect to \eqref{e:deta}. However, during the experiments they were chosen so that virtual inertia is several times larger than inertia of comparable SG.
\subsection{Voltage control}
Idea behind voltage control is significantly simpler. It here proportional integral control is used similar to standard excitation systems:
\begin{subequations}\label{e:vc}
\begin{equation}
    V_{set}(t) = c_P(V_{ref}(t)-V_{out}(t))+c_I\int_0^t(V_{ref}(\tau)-V_{out}(\tau))d\tau.
\end{equation}
\end{subequations}

\section{Zero-inertia grid setup}\label{s:setup}
\begin{figure}
    \centering
    \includegraphics[width=1\linewidth]{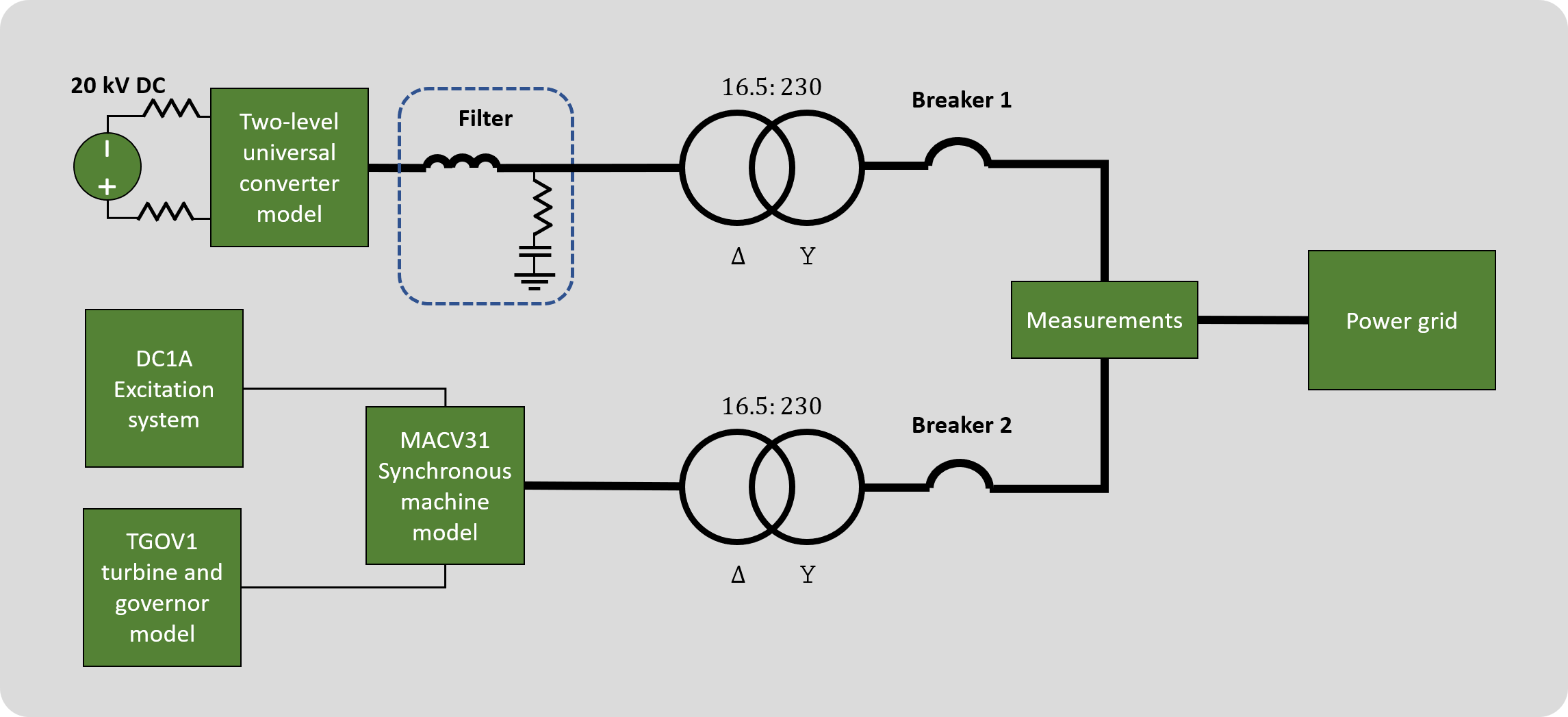}
    \caption{Generating unit setup}
    \label{fig:gu}
\end{figure}

The modified IEEE 9-bus system was utilized for investigation \cite{Sauer1998}. System's generation and load parameters are presented in Table \ref{tab:ps}. The table lists the capacities of inverters used as alternatives to synchronous generation. The system was tested in its traditional form with three SGs and in a modified form with an IBRs. Loads were connected to buses 5, 6, and 8. Key parameters of the 9-bus system, including line lengths, primary load capacities, and SG capacities, remained consistent. One test scenario involved adding an additional 30 MW of power to the buses with loads to observe the expected frequency drop with fully synchronous generation.  In order to simulate different system setups, each generating unit is equipped with both SG and IBR (Fig. \ref{fig:gu}). Parameters of the generating units are given in Table \ref{tab:gu}. Here SG damping is calculated according to the formula \cite{MBB}
\begin{equation}
    D = \frac{X'_q - X''_q}{X+X'_q}\frac{X'_q}{X''_q}T''_qV_{out},
\end{equation}
where $X$ is step-up transformer reactance, $T''_q$ is open circuit q-axis time constant,  $X'_q$ and $X''_q$ are q-axis transient and subtransient reactances respectively.
\begin{table}[]
    \centering
    \caption{Inertia and Dumping coefficients}
    \begin{tabular}{|c|c|c|}\hline
         & SG & Inverter \\\hline
        Inertia (MWs/MVAR) & $23.64$ & $25-100$ \\
        Dumping (MW/rad) & $68.95$ & $50-200$ \\\hline
    \end{tabular}
    \label{tab:gu}
\end{table}

\begin{figure}[b!]
    \centerline{\includegraphics[width=0.47\textwidth]{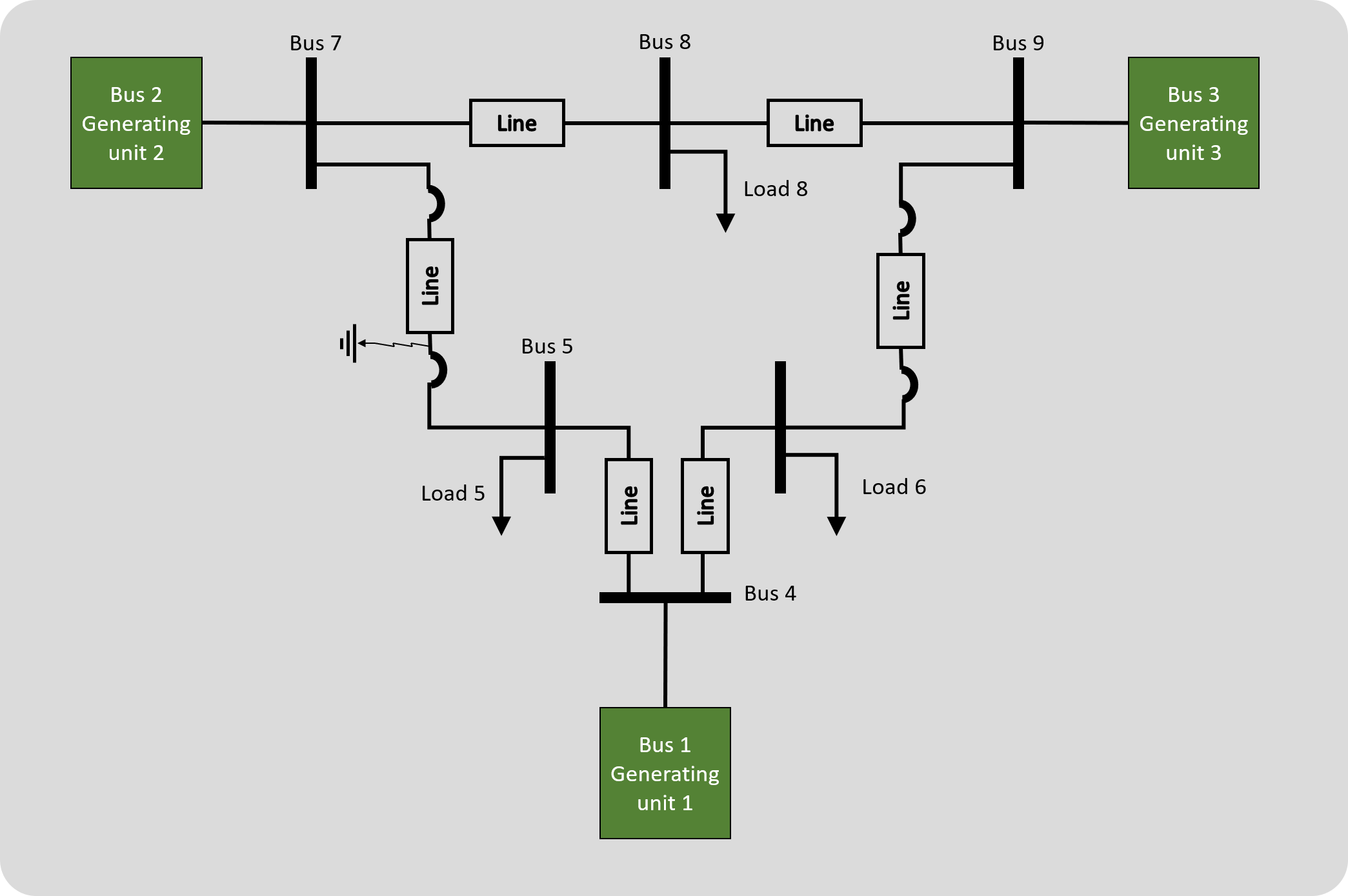}}
    \caption{9-bus power system in RSCAD.}
    \label{fig:ninebus}
\end{figure}

Modeling was conducted using the Novacor 1.0 RTDS. The RSCAD model is illustrated in Fig. \ref{fig:ninebus}. The detailed inverter component is modeled using a substep mode with 2-5 microsecond timestep, employing a two-level substep bridge and a universal two-level converter model. The control system of the inverter immediately adopts a GFM type, integrating the Pontryagin's Maximum Principle with the VSG concept. The operation of the switches in each bridge is managed by a firing pulse generator from the RSCAD user library, which receives modulation wave inputs. The transition from the substep to the main timestep mode is carried out using substep interface transformers.

\begin{table}[b!] 
\centering
\caption{Inverter-based IEEE 9-Bus System Parameters}
\label{tab:ps}
\setlength{\tabcolsep}{3pt}
\begin{tabular}{|p{25pt}|p{45pt}|p{45pt}|p{45pt}|p{45pt}|}
\hline
\textbf{Bus} & \textbf{IBR capacity (MW)} & \textbf{IBR capacity (MVar)} & \textbf{Load (MW)} & \textbf{Load (MVar)} \\
\hline
1 & 150 & 200 & - & - \\
2 & 250 & 300 & - & - \\
3 & 100 & 200 & - & - \\
4 & - & - & - & - \\
5 & - & - & 125 + 30 & 50 + 0.01 \\
6 & - & - & 90 + 30 & 30 + 0.01 \\
7 & - & - & - & - \\
8 & - & - & 100 + 30 & 35 + 0.01 \\
9 & - & - & - & - \\
\hline
\end{tabular}
\end{table}

\section{System resilience tests}\label{s:tests}
All tests aimed to study the network operation features with partial and complete replacement of synchronous generation with an IBR. The test scenarios included:

\begin{enumerate}
  \item Fault on line 5-7, line shutdown to eliminate the fault, and auto-reclosure of the line.
  \item Adding additional power to the load buses.
  \item Island mode operation of network sections when lines 5-7 and 6-9 are disconnected.
\end{enumerate}

Experimental conditions:

\begin{enumerate}
  \item All generation is synchronous.
  \item One IBR, two SGs.
  \item Two IBRs, one SG.
  \item All IBRs.
\end{enumerate}

Assumptions:

\begin{enumerate}
  \item The fault type simulated is a severe three-phase to ground fault.
  \item The functionality of protection and automation systems within the electric power system during faults is considered without detailed modeling of the underlying algorithms.
  \item To observe the frequency sag effect, underfrequency load shedding is not simulated.
  \item Two fault duration times on line 5-7 are simulated: 15 milliseconds (assuming correct high-speed line protection operation) and 300 milliseconds (possible in emergencies like failure of the main protection stage).
  \item Observed system parameters: frequency, voltage, active and reactive power.
\end{enumerate}

\subsection{Load connection}
\begin{figure}[b!]
    \centerline{\includegraphics[width=0.3\textwidth]{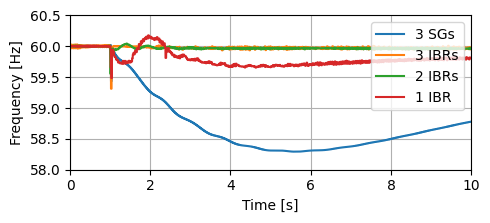}}
    \caption{Frequencies during additional load integration.}
    \label{fig:load_freq}
\end{figure}

Consider a scenario in which an additional 30 MW load is connected to each bus. Fig. \ref{fig:load_freq} illustrates the resulting frequencies within the network. To highlight the response time of inverter-based GFM sources, a 10-second time window was selected. The frequency in a conventional 9-bus system equipped with three SGs (depicted by the blue curve) experiences a significant drop upon connection and does not recover. Conversely, as the share of IBRs in the network increases (represented by the red, green, and orange curves), the frequency recovery rate improves.

\begin{figure}[b!]
    \centerline{\includegraphics[width=0.3\textwidth]{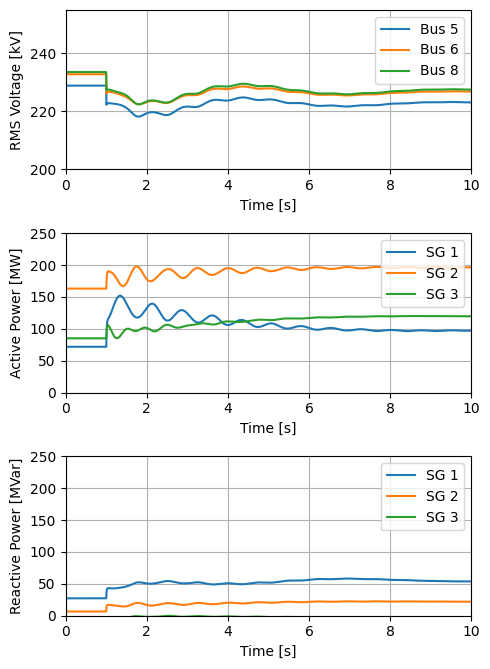}}
    \caption{Operating parameters under fully synchronous generation during additional load integration.}
    \label{fig:load_vpq_sg}
\end{figure}

\begin{figure}[b!]
    \centerline{\includegraphics[width=0.3\textwidth]{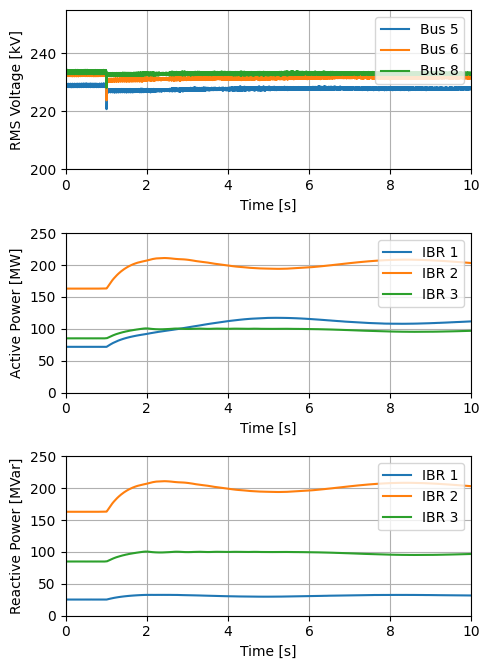}}
    \caption{Operating parameters under fully inverter-based generation during additional load integration.}
    \label{fig:load_vpq_3inv}
\end{figure}

Fig. \ref{fig:load_vpq_sg} and \ref{fig:load_vpq_3inv} present the RMS voltages, as well as the active and reactive power graphs. When compared to the initial scenario depicted in Fig. \ref{fig:load_vpq_sg}, which involves fully synchronous generation, the installation of three IBRs results in the voltages on the load buses remaining nearly constant. Additionally, the variations in the injected powers become smoother.

\subsection{Fault on the line 5-7}

\begin{figure}[b!]
    \centerline{\includegraphics[width=0.3\textwidth]{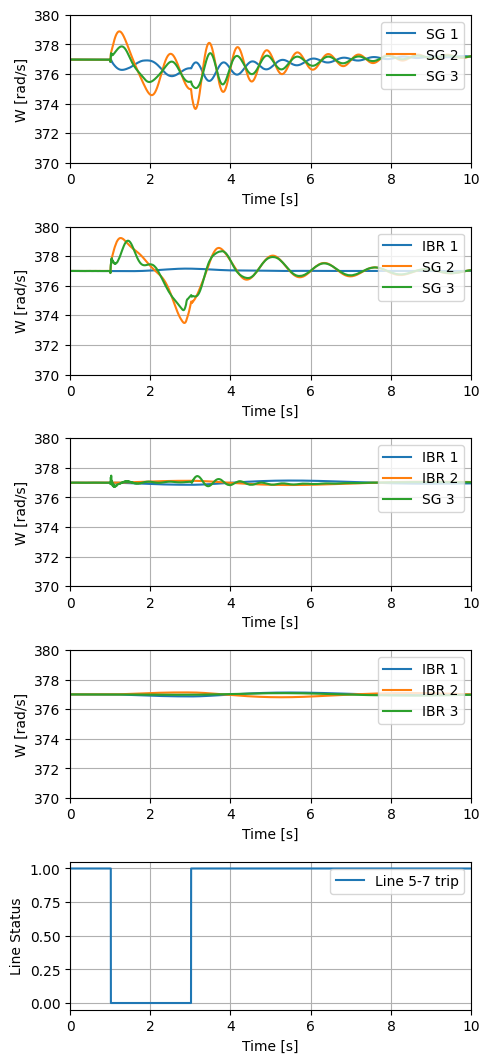}}
    \caption{Rotational speed at different generating units.}
    \label{fig:flt_w}
\end{figure}

Next, we examine a scenario involving a 15-millisecond fault on line 5-7. As shown in Fig. \ref{fig:flt_w}, the rotational speed is better maintained with an increased share of IBR in the network, even during an emergency. While the network remains stable in all cases, the rotational speed in a fully synchronous generation scenario stabilizes in approximately 9 seconds. In contrast, with fully inverter-based generation, the rotational speed remains unaffected. Furthermore, it is noteworthy that an increased share of IBR positively influences the SGs operating in parallel with the IBRs. The transient mode duration is reduced, and the rotational speed recovers more rapidly.

\begin{figure}[b!]
    \centerline{\includegraphics[width=0.3\textwidth]{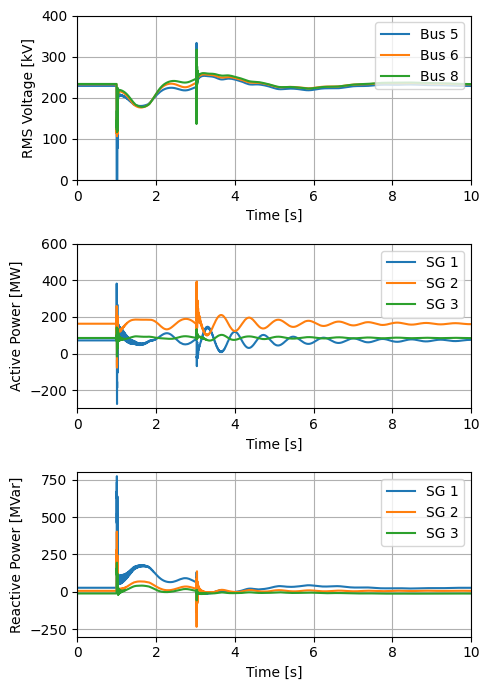}}
    \caption{Fully synchronous generation and operating parameters at fault on the line.}
    \label{fig:flt_3sg}
\end{figure}

\begin{figure}[b!]
    \centerline{\includegraphics[width=0.3\textwidth]{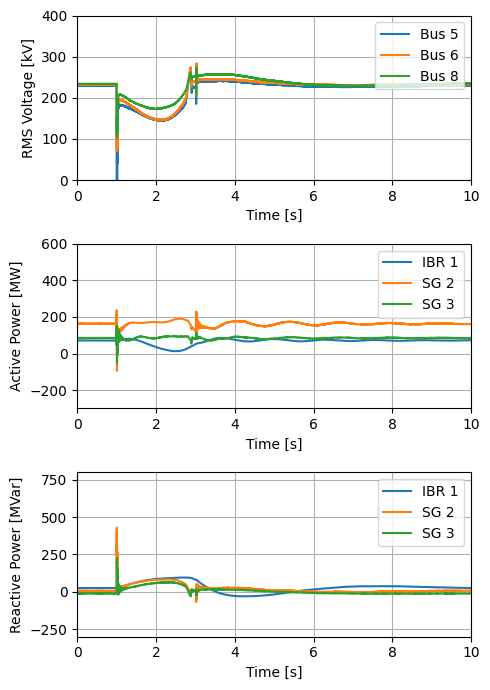}}
    \caption{One IBR, two SGs, and operating parameters at fault on the line.}
    \label{fig:flt_1inv}
\end{figure}

\begin{figure}[b!]
    \centerline{\includegraphics[width=0.3\textwidth]{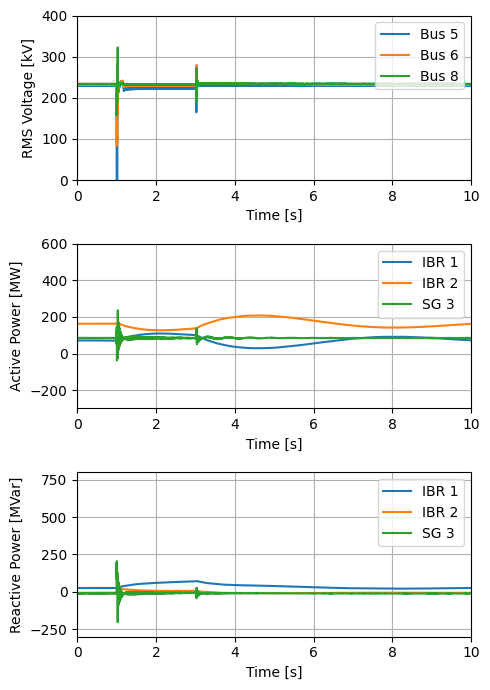}}
    \caption{Two IBRs, one SG, and operating parameters at fault on the line.}
    \label{fig:flt_2inv}
\end{figure}

\begin{figure}[b!]
    \centerline{\includegraphics[width=0.3\textwidth]{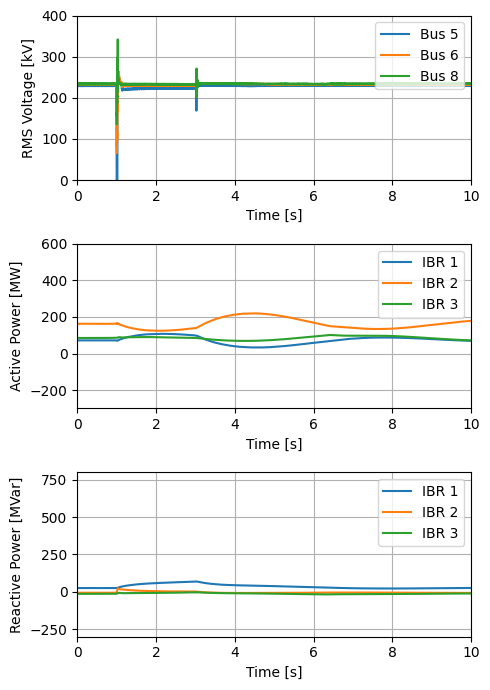}}
    \caption{Fully inverter generation and operating parameters at fault on the line.}
    \label{fig:flt_3inv}
\end{figure}

Fig. \ref{fig:flt_3sg}-\ref{fig:flt_3inv} demonstrate the operating parameters as synchronous generation is sequentially replaced with IBR. There is a clear trend toward smoother transient processes and reduced recovery times for the operating parameters.

In a network configuration with one IBR and two SGs (Fig. \ref{fig:flt_1inv}), voltage recovery is faster compared to the fully synchronous generation scenario (Fig. \ref{fig:flt_3sg}). With two and three IBRs (Fig. \ref{fig:flt_2inv} and \ref{fig:flt_3inv}), the voltage remains nearly constant.

Fig. \ref{fig:flt_3inv} also reveals that the nature of power surges during and after the fault becomes smoother. Active power remains within 250 MW, and reactive power within 80 MVAr. In contrast, the active power surge in Fig. \ref{fig:flt_3sg} reaches up to 400 MW, with reactive power spikes up to 800 MVAr. In Fig. \ref{fig:flt_1inv} and \ref{fig:flt_2inv}, power surges are somewhat reduced. Such power surges, as described, have the potential to damage electrical equipment.

\begin{figure}[b!]
    \centerline{\includegraphics[width=0.3\textwidth]{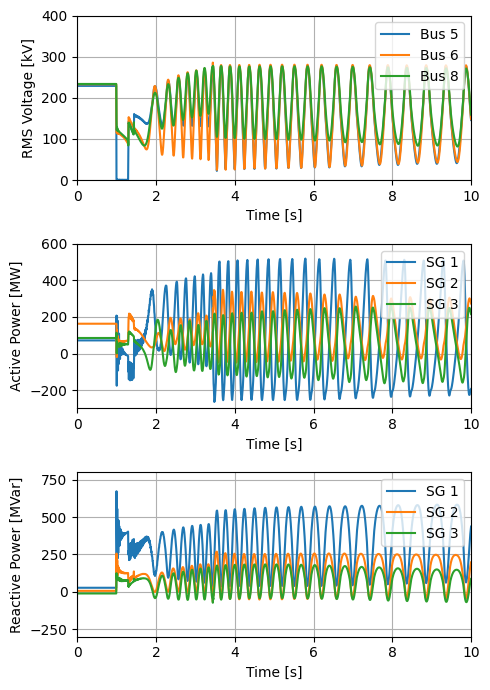}}
    \caption{Operating parameters at long fault with fully synchronous generation.}
    \label{fig:flt03_3sg_vpq}
\end{figure}

Consider a scenario involving the abnormal operation of the protection system for line 5-7. Similar to previous cases, a fault occurs, the line shuts down, and then reclosure happens after the fault is cleared. In this instance, the fault duration is 300 milliseconds. Fig. \ref{fig:flt03_3sg_vpq} illustrates a scenario where a prolonged fault occurs in a network with fully synchronous generation. Following the fault, oscillations with an amplitude of nearly 200 kV commence. Grid fails to survive; in reality, this would result in a blackout and significant damage to the infrastructure.

\begin{figure}[t!]
    \centerline{\includegraphics[width=0.3\textwidth]{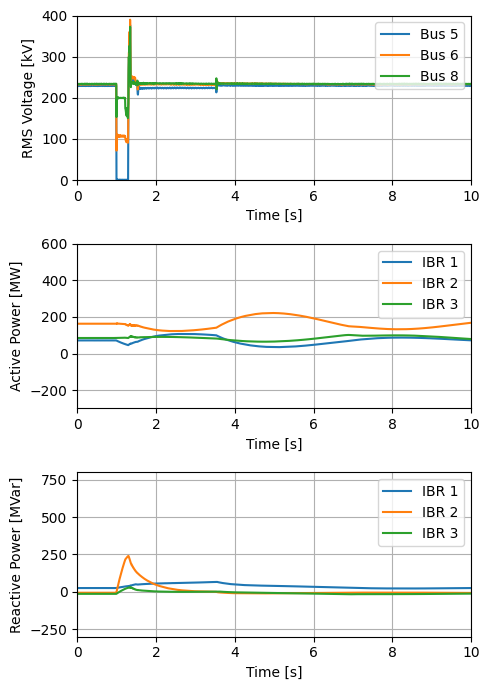}}
    \caption{Operating parameters at long fault with fully inverter generation.}
    \label{fig:flt03_3inv_vpq}
\end{figure}

In contrast, Fig. \ref{fig:flt03_3inv_vpq} presents a scenario where a prolonged fault occurs in a network with fully inverter-based generation. Despite a brief voltage surge reaching up to 400 kV and a reactive power surge up to 250 MVAr once the fault is cleared, the network remains intact. This experiment demonstrates the comparative stability of an inverter-based power supply network under prolonged emergency conditions.

\subsection{Generating unit disconnection}

To further assess the reliability of the network, it is important to consider the scenario of disconnecting one of the generating units. Two cases were tested for representativeness: one with fully synchronous generation and the other with fully inverter-based generation, with the first generating unit disconnected.

\begin{figure}[b!]
    \centerline{\includegraphics[width=0.3\textwidth]{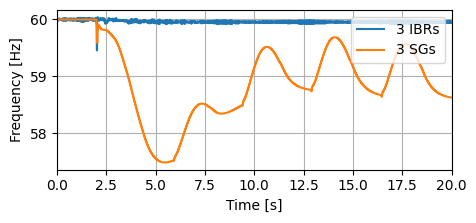}}
    \caption{Frequencies at disconnection of one generating unit in two modes.}
    \label{fig:disc_freq}
\end{figure}

Fig. \ref{fig:disc_freq} shows the frequency oscillograms over a selected 20-second period for clarity. When one SG is disconnected (orange curve), the frequency drops to 57.5 Hz, which is critically low for the power system, and it does not recover, continuing to oscillate. This would lead to a system collapse in reality. Conversely, when one IBR is disconnected (blue curve), the frequency drops to 59.5 Hz but quickly recovers to its initial value.

\begin{figure}[t!]
    \centerline{\includegraphics[width=0.3\textwidth]{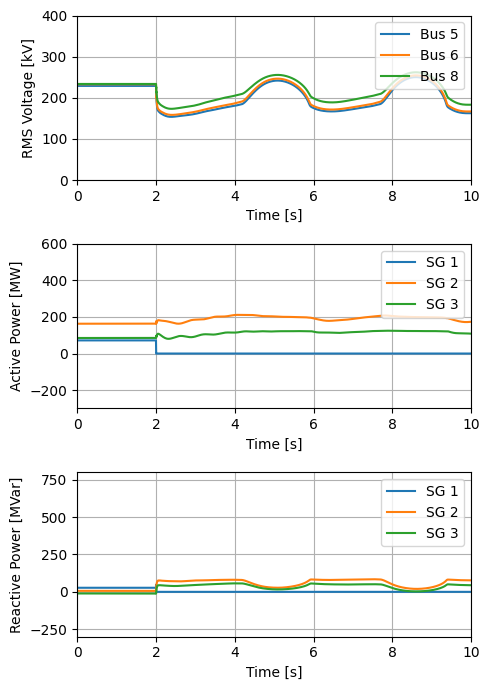}}
    \caption{Operating parameters at disconnection of one generating unit with fully synchronous generation.}
    \label{fig:disc_1sg_vpq}
\end{figure}

Fig. \ref{fig:disc_1sg_vpq} depicts the operating parameters in the case of fully synchronous generation. It shows a voltage drop to nearly 160 kV, and the power levels are unstable.

\begin{figure}[t!]
    \centerline{\includegraphics[width=0.3\textwidth]{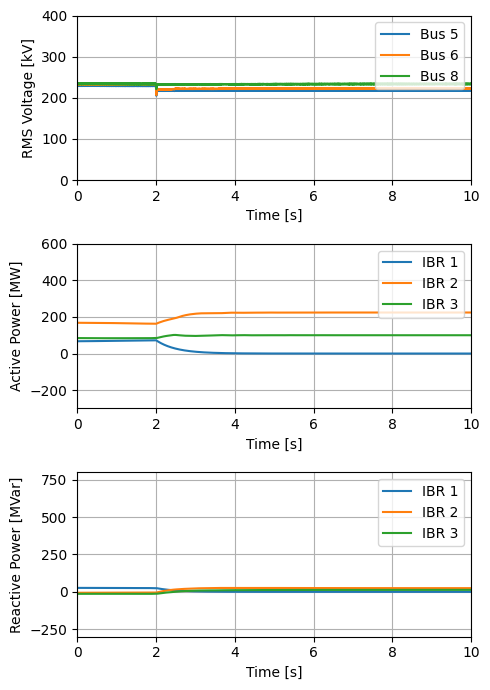}}
    \caption{Operating parameters at disconnection of one generating unit with fully inverter generation.}
    \label{fig:disc_1inv_vpq}
\end{figure}

Fig. \ref{fig:disc_1inv_vpq} illustrates the case of fully inverter-based generation, highlighting the differences in network stability.

\subsection{Island mode}

\begin{figure}[t!]
    \centerline{\includegraphics[width=0.3\textwidth]{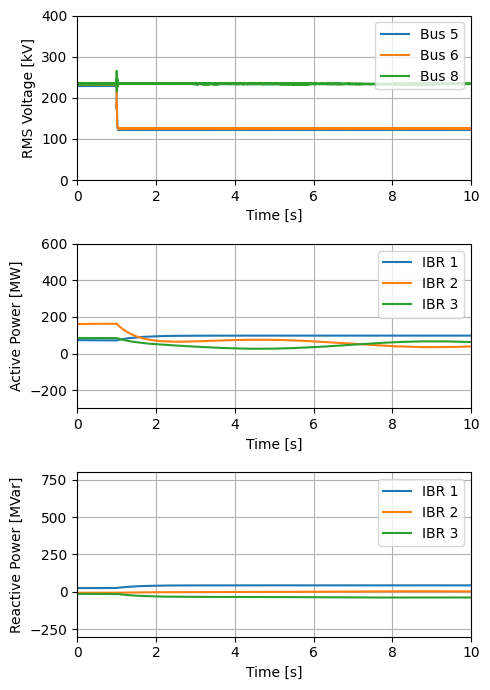}}
    \caption{Operating parameters in island mode with fully inverter generation.}
    \label{fig:island_inv}
\end{figure}

We also examine a scenario with fully inverter-based generation operating in island mode, which occurs when lines 5-7 and 6-9 are disconnected, possibly due to protection system failures or accidents. In this mode, inverters 2 and 3 supply the load on bus 8, while inverter 1 attempts to distribute power among the loads on buses 5 and 6. Fig. \ref{fig:island_inv} shows that the voltage for the load on bus 8 remains stable and nearly constant. The capacities of inverters 2 and 3 are effectively distributed. However, the situation for the loads on buses 5 and 6 is different; the voltage drops to around 120 kV. Nevertheless, the network survives and continues to function in a new steady-state mode. The issue of insufficient voltage could be addressed by installing a more powerful inverter.

\section{Conclusions}\label{s:conclusion}
This paper investigates dynamics of low- and zero inertia grids with GFM inverters. Developed inverter control represents a VSG with dynamic dumping and inertia together with additional shift functions for inverter's frequency and phase angle. The developed approach allows to keep virtual system inertia up to 4 times large and dumping up to 3 times large than the corresponding parameters of comparable SGs. Dynamic stability of the developed control approach is theoretically proven. Theoretical results are supported by extensive case study in RTDS with detailed two-level universal converter model operating at 1-3 microseconds timestep. The studies include large step-changes in loads, islanding mode and short circuits events for IEEE 9-bus case. In all experiments inverters with the developed control demonstrate superior dynamics performance and increased robustness in comparison to stadard SGs.

\bibliographystyle{IEEEtran}
\bibliography{main.bib}
\begin{IEEEbiography}[{\includegraphics[width=1in,height=1.25in,clip,keepaspectratio]{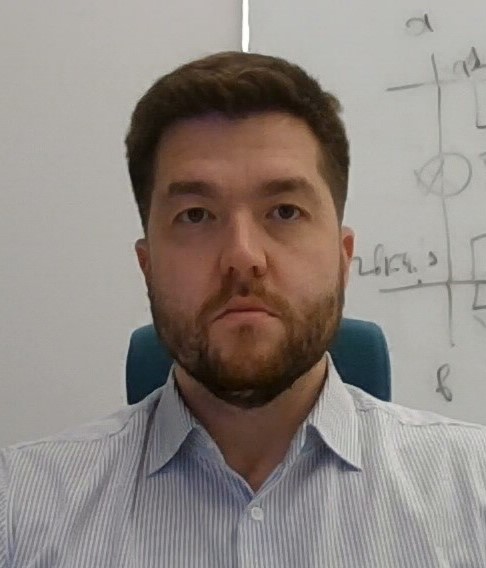}}]{Oleg O. Khamisov}
received M.Sc. in Applied Mathematics from Irkutsk State University, Russia in 2015 and Ph.D. in Engineering Systems from Skolkovo Institute of Science and Technology (Skoltech), Russia in 2020. Currently he is an assistant professor at Skoltech. His research is focused on development of control algorithms  for low-inertia systems.
\end{IEEEbiography}
\begin{IEEEbiography}[{\includegraphics[width=1in,height=1.25in,clip,keepaspectratio]{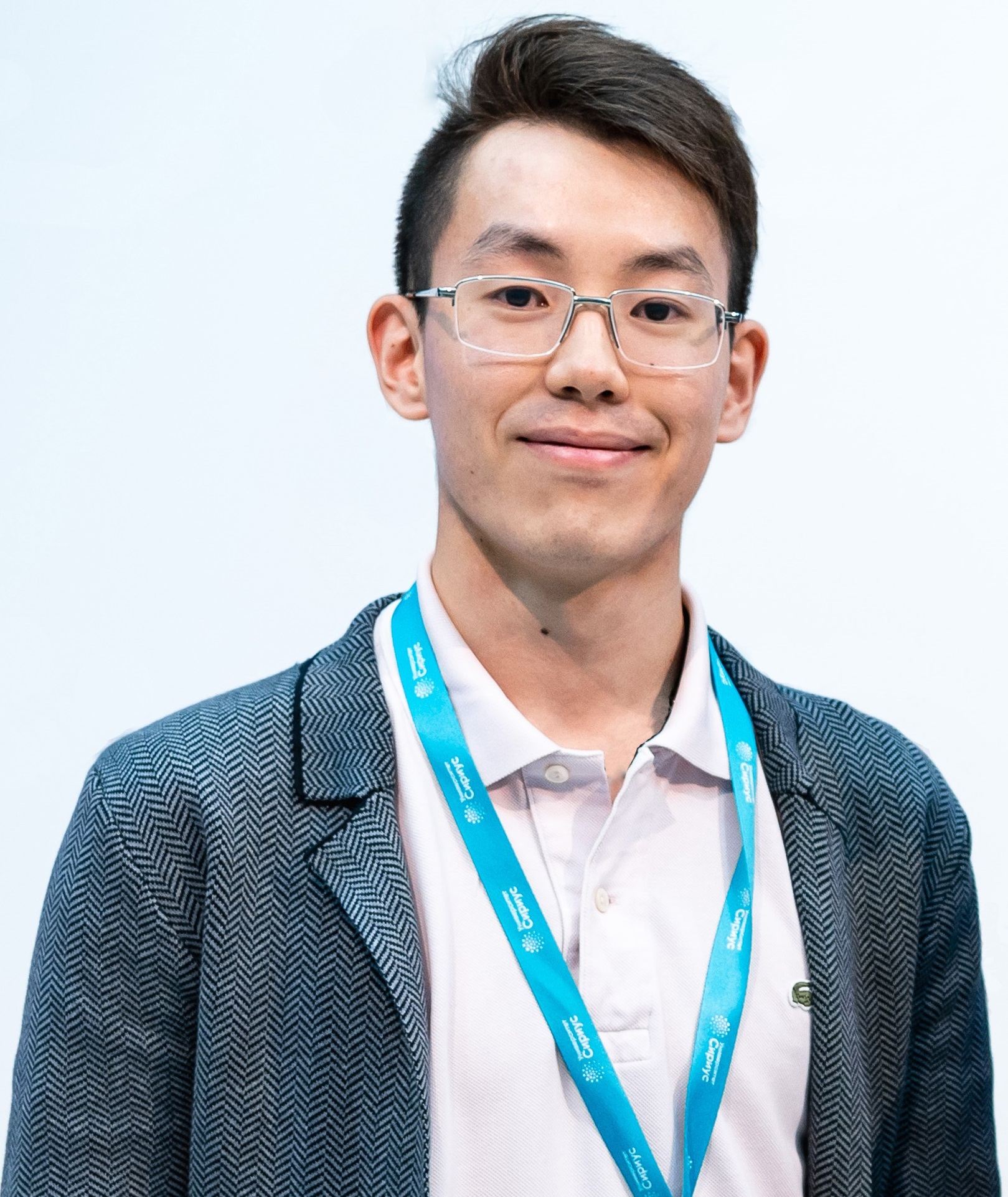}}]{Stepan P. Vasilev} received M.Sc. in Intelligent Protection, Automation and Control for Distributed Energy Systems from the Moscow Power Engineering Institute, Russia, in 2020. He is currently pursuing the Ph.D. degree with the Skolkovo Institute of Science and Technology. His research interest includes applied machine learning, power system comprehensive simulations and stability analysis.
\end{IEEEbiography}
\end{document}